\documentclass[12pt]{iopart}

\usepackage{graphicx}
\usepackage{colordvi}
\begin{document}
\def\H{\,{\mathcal  H}}
\def\bea{\begin{eqnarray}}
\def\eea{\end{eqnarray}}

\title{Signatures of Primordial Non-Gaussianity in
 the Large-Scale Structure of the 
Universe }
\author{Nicola Bartolo $^{(1)}$, Sabino
Matarrese $^{(2,3)}$ and Antonio Riotto$^{(3)}$}
\address{$^{(1)}${\it The Abdus Salam International Centre for Theoretical 
Physics, Strada Costiera 11, 34100, Trieste, Italy}}
\address{$^{(2)}${\it Dipartimento di Fisica `Galileo Galilei', 
Universit\`a di Padova, via Marzolo 8, I-35131, Padova, 
Italy}}
\address{$^{(3)}${\it INFN, Sezione di Padova,
via Marzolo 8, I-35131, Padova, Italy}}

\date{\today}

\begin{abstract}
\noindent
We discuss how primordial ({\it e.g.} inflationary) non-Gaussianity in the 
cosmological 
perturbations is left imprinted in the Large-Scale Structure  of the universe. 
Our findings show  that the information on the
primordial non-Gaussianity 
 set on super-Hubble scales flows into Post-Newtonian  terms, leaving
an observable imprint in the Large-Scale Structure. 
Future high-precision measurements
of the statistics of  the dark matter density and peculiar 
velocity fields will allow to pin down the primordial non-Gaussianity,
thus representing a tool complementary to studies of the Cosmic
Microwave Background anisotropies. 
\noindent

\end{abstract}


\maketitle

\noindent
The most compelling  feature  of inflation \cite{lrreview} is to
provide the seeds for the Large-Scale Structure (LSS) of the universe
and the anisotropies in the Cosmic Microwave Background.
Cosmological  perturbations are created during inflation from
quantum fluctuations and ``redshifted'' to sizes larger than
the Hubble radius. The deviation from a Gaussian statistics
characterizing the cosmological perturbations, {\it i.e.} the
presence of higher-order connected correlation functions,  
emerges as a key observable 
to discriminate among the various scenarios for the 
generation of cosmological perturbations produced during inflation.
Non-Gaussianity (NG)  
is one of the primary targets of present and future 
Cosmic Microwave Background (CMB) satellite missions \cite{review}.

The main goals of this {\it Letter} are to show how   
NG -- set at primordial epochs on 
large (super-Hubble) scales --
propagates to subhorizon scales once 
cosmological perturbations  reenter the Hubble radius 
during the matter- and dark energy-dominated epochs
and to analyze which  signatures  primordial 
NG leaves in the Large-Scale Structure (LSS) of the
universe through the dark matter density and velocity
fields. 
We start with considering the case of 
flat universe filled by a perfect  
non-relativistic matter fluid. For the purposes of this paper this is a good 
approximation, although for the scales crossing the Hubble radius during 
the radiation dominated era one should include the radiation in the evolution 
equations, thus using a complete transfer function also for those scales.
We find it convenient to perform our calculations in the comoving-synchronous 
gauge and then switch to the Poisson gauge~\cite{Bert}, which 
allows a more direct comparison with the standard Newtonian and Eulerian 
approach
adopted in LSS studies. 
We make use of the  
formalism developed in Refs.~\cite{MT,MMB} which the reader is referred to
for more details.   
The synchronous-gauge is defined by the line-element $
ds^2=a^2(\tau)[-d\tau^2+\gamma_{ij}({\bf x},\tau) dx^i dx^j]$, 
where $a(\tau)$ is the scale factor, $\tau$ is the conformal time 
and ${\bf x}$ represent  
comoving Lagrangian coordinates for the fluid element. 
The energy-momentum 
tensor is $T^{\mu \nu}=\rho u^{\mu} u^{\nu}$, where $\rho$ is the mass
density and, in our comoving coordinates, the fluid 
four-velocity is given by $u^{\mu}=(1/a,0,0,0)$. The homogeneous background 
(Einstein-de Sitter universe) is described by a scale factor 
$a(\tau)\propto\tau^2$. 
A very efficient way to write down  Einstein and continuity equations is 
to introduce the peculiar velocity-gradient~\cite{MT} 
$\vartheta^{i}_j\equiv 
u^{i}_{~;j}-\frac{a'}{a}\delta^{i}_{j}=\frac{1}{2}\gamma^{ia}
\gamma'_{aj}$, where we have subtracted the isotropic Hubble flow. 
Primes stand for differentiation w.r.t. the conformal time; 
semicolons denote  covariant 
differentation. From 
the continuity equation $T^{\mu\nu}_{~~~;\nu}=0$, we infer  
the  exact solution for the density contrast $\delta=\delta \rho/\rho$~
\cite{MT,MMB}
\label{delta}   
\begin{equation}
\label{delta}
\delta({\bf x, \tau})=(1+\delta_{\rm in}({\bf x}))[\gamma({\bf x},\tau)
/\gamma_{\rm in}({\bf x})]^{-1/2}-1\, ,
\end{equation}
where $\gamma={\rm det} \gamma_{ij}$. The subscript ``${\rm in}$'' denotes the 
value of quantities evaluated 
at some initial time.  Such initial 
conditions will 
play a key role once they  have been   properly 
determined to account for the  primordial non-linearities in the 
cosmological perturbations. 
 
From Eq.~(\ref{delta}) it is evident that in the comoving-synchronous
gauge the real independent 
degree of freedom is the spatial metric tensor $\gamma_{ij}$.
The energy constraint reads
\begin{equation}
\label{energycon}
\vartheta^2-\vartheta^i_{~j}\vartheta^j_{~i}+\frac{8}{\tau}\vartheta
+{\mathcal R}=
\frac{24}{\tau^2} \delta \, ,
\end{equation}  
where ${\mathcal R}^i_{~j}$ is the Ricci tensor associated with the spatial 
metric $\gamma_{ij}$ with scalar curvature ${\mathcal R}={\mathcal R}^i_{~i}$.
The momentum constraint reads
$\vartheta^i_{~j|i}=\vartheta_{,j}$;  bars stand for a covariant 
differentiation  in the three-space with metric
$\gamma_{ij}$. Finally, the Raychaudhuri equation 
\begin{equation}
\label{Ray}
\vartheta'+\frac{2}{\tau}\vartheta+\vartheta^i_{~j}\vartheta^j_{~i}+
\frac{6}{\tau^2}\delta=0\, ,
\end{equation} 
is obtained from the energy constraint and the trace of the 
evolution equation
 $\vartheta^{i'}_{~j}+4\vartheta^i_{~j}/\tau
+\vartheta\vartheta^{i}_{~j}+(\vartheta^k_{~l} \vartheta^l_{~k}-\vartheta^2)
\delta^i_{~j}/4+{\mathcal R}^i_{~j}
-{\mathcal R} \delta^i_{~j}/4=0$. 
Notice that these equations are exact 
and describe the fully non-linear evolution of cosmological perturbations 
around the Einstein-de Sitter universe (up to caustic formation). 
In order to show how the
primordial  non-Gaussianities show up in 
 in the matter density contrast,  we  make a 
perturbative expansion up to second order in the fluctuations of the 
metric and matter variables. The spatial metric tensor can  be written as 
$\gamma_{ij}=(1-2\psi^{(1)}-\psi^{(2)})\, \delta_{ij}
+\chi^{(1)}_{ij}+\frac{1}{2} 
\chi^{(2)}_{ij}$, 
where $\chi^{(1)}_{ij}$ and $\chi^{(2)}_{ij}$ are 
traceless and include scalar, vector and tensor (gravitational waves) 
perturbations. Similarly,  we split 
the density contrast into a linear and a second-order part as 
$\delta({\bf x},\tau)=\delta^{(1)}({\bf x},\tau)+\frac{1}{2}
\delta^{(2)}({\bf x},\tau)$. At linear order the growing-mode solutions 
for a matter-dominated epoch in the comoving-synchronous gauge are given by
~\cite{MMB} 
$\psi^{(1)}({\bf x},\tau)=\frac{5}{3}\varphi({\bf x})+\frac{\tau^2}{18} 
\nabla^2\varphi({\bf x})$, and 
$\chi^{(1)}_{ij}=-\frac{\tau^2}{3}\left(\varphi_{,ij}-\frac{1}{3}\delta_{ij}
\nabla^2\varphi \right)$ for the metric perturbations and
$\delta^{(1)}=\frac{\tau^2}{6} \nabla^2\varphi$ 
for the linear density contrast. Here $\varphi({\bf x})$ is 
the so-called~{\it peculiar gravitational potential}. 
In writing $\chi^{(1)}_{ij}$ 
we have eliminated the residual gauge ambiguity of the synchronous gauge as
in Ref.~\cite{MMB} and 
we have neglected linear vector modes since they are not produced in standard 
mechanisms for the generation of cosmological perturbations (as inflation). 
We have also neglected linear tensor modes, since they play a negligible
role in LSS formation.

Let us now discuss the key issue of the  initial 
conditions which we conveniently 
fix at the   time when the cosmological perturbations
relevant today for LSS are well outside the Hubble radius, 
{\emph i.e.} when the (comoving) wavelength $\lambda$ 
of a given perturbation mode is such that
 $\lambda \gg {\mathcal H}^{-1}$,  
${\mathcal H=a'/a}$ being the conformal Hubble constant. 

In the standard single-field inflationary model, 
the first seeds of density fluctuations are laid down on super-horizon scales 
from the fluctuations of a scalar field, the inflaton \cite{lrreview}. 
Recently many other scenarios have been proposed as alternative 
mechanisms to generate such primordial seeds. 
These include, for example, the curvaton~\cite{curvaton} and the 
inhomogeneous reheating scenarios~\cite{gamma1}, 
where the first density 
fluctuations are produced through the fluctuations of a scalar field
other than the inflaton. 
In order to follow the evolution on super-Hubble scales of the  
density fluctuations coming from the various  mechanisms, we 
use the curvature 
perturbation $\zeta=\zeta^{(1)}+(1/2)\zeta^{(2)}+\cdots$
of uniform density hypersurfaces, where 
$\zeta^{(1)}=-\psi^{(1)}-{\mathcal H} \frac{\delta^{(1)} \rho}{\rho'}$ and
the expression for $ \zeta^{(2)}$ can be found in  
Ref.~\cite{mw}.  
The crucial point is that 
the  gauge-invariant curvature perturbation
$\zeta$ remains  {\it constant} on super-Hubble scales after it 
has been generated duirng a primordial epoch and possible isocurvature 
perturbations are no longer present. Therefore, we may set
the initial conditions for LSS at the time when $\zeta$  becomes
constant. In particular,   $\zeta^{(2)}$ 
provides  the necessary information about the
``primordial'' level of NG generated either during inflation, as in the 
standard scenario, or immediately after it, as in the curvaton scenario. 

Different scenarios are  characterized by different values of 
$\zeta^{(2)}$. For example, in    
the standard single-field inflationary model
 $\zeta^{(2)}=2\left( 
\zeta^{(1)} \right)^2+{\cal O}\left(n_\zeta-1\right) $~\cite{ABMR,BMR2}, where 
$n_\zeta$ is the spectral index for the scalar perturbations. 
In general, we may  parametrize the primordial NG level 
in terms of the conserved curvature 
perturbation  as in Ref. \cite{Tinv},  
$\zeta^{(2)}=2a_{\rm nl}\left(\zeta^{(1)}\right)^2$.
In the standard scenario $a_{\rm nl}\simeq 1$, in the  
curvaton case $a_{\rm nl}=(3/4r)-r/2$, where 
$r \approx (\rho_\sigma/\rho)_{\rm D}$ is the relative   
curvaton contribution to the total energy density at curvaton 
decay~\cite{review}. In the minimal picture for the inhomogeneous 
reheating scenario, $a_{\rm nl}=1/4$. For other scenarios we refer 
the reader to Ref.~\cite{review}. 
The curvature perturbation $\zeta=\zeta_{\rm in}$ allows to  set the initial 
conditions for the metric and matter perturbations accounting for the 
primordial contributions. 
Indeed,  in the synchronous gauge and on super-Hubble scales 
the energy density contrast 
vanishes (see, {\emph e.g.}, Ref.~\cite{Kolbetal}) 
and therefore $\delta_{\rm in}=0$. From the definiton of $\zeta^{(1)}$ 
it then follows that $\zeta^{(1)}\simeq -\psi^{(1)}=-5 \varphi/3$ and at 
second-order 
\begin{equation}
\zeta^{(2)}\simeq-\psi^{(2)}_{\rm in}=2 a_{\rm nl} \left(\zeta_{\rm in}^{(1)}
\right)^2=\frac{50}{9}\,a_{\rm nl}\, \varphi^2\, ,
\end{equation} 
where $a_{\rm nl}$ will  always signal the presence of primordial NG 
according to our parametrization. One of the best tools 
to detect or constrain the primordial large-scale NG is through the analysis 
of the CMB anisotropies, for example by studying the bispectrum
~\cite{review}. In that case the standard procedure is to
introduce  the non-linearity 
parameter $f_{\rm nl}$ characterizing NG in the large-scale 
temperature anisotropies~\cite{ks,k,review}. To give the feeling
of the resulting size of $f_{\rm nl}$ when $|a_{\rm nl}| \gg 1$, 
$f_{\rm nl} 
\simeq -5 a_{\rm nl}/3$~(see Refs.~\cite{review,Tinv}, where the sign is 
opposite for a different 
convention relating the CMB anisotropies to the gravitational potential). 

After having learned how to set initial conditions, our next 
step is to 
determine how a primordial NG
evolves onto subhorizon scales. By perturbing Eq.~(\ref{delta}) 
up to second order, 
we get 
\begin{eqnarray}
\label{delta2s}
\delta&=&3(\psi^{(1)}-\psi^{(1)}_{\rm in})+\delta^{(1)}_{\rm in}
+\frac{3}{2}(\psi^{(2)}-\psi^{(2)}_{\rm in})+\frac{1}{2}
\delta^{(2)}_{\rm in}
+\frac{1}{8}(\gamma^{(1)j}_j-\gamma^{(1)j}_{{\rm in}~j})^2
\nonumber \\
&+&\frac{1}{4}(\gamma^{(1)ij}\gamma^{(1)}_{ij}
-\gamma^{(1)jk}_{\rm in}\gamma^{(1)}_{{\rm in}~jk})
-\frac{1}{2}\delta^{(1)}_{\rm in}(\gamma^{(1)j}_j-
\gamma^{(1)j}_{{\rm in}~j})\, ,
\end{eqnarray}
where $\gamma_{ij}\equiv \delta_{ij}+\gamma^{(1)}_{ij}+\gamma^{(2)}_{ij}/2$.
To compute the metric perturbation $\psi^{(2)}$, we need 
only the Raychaudhury 
equation~(\ref{Ray}) 
and the energy constraint~(\ref{energycon}). Perturbing these equations up 
to second oder and inserting the solution for the density 
contrast~(\ref{delta2s}) in terms of the spatial metric perturbations, one 
obtains an expression for $\psi^{(2)}$ in terms of the peculiar gravitational 
potential $\varphi({\bf x})$     
\begin{eqnarray}
\label{psi2s}
\psi^{(2)}&=&-\frac{50}{9}a_{\rm nl} \varphi^2+\frac{10}{27} 
\left( \frac{3}{4}-a_{\rm nl}  \right) \tau^2 \varphi^{,k}\varphi_{,k} 
+\frac{10}{27} (1-a_{\rm nl} ) \tau^2 \varphi \nabla^2 \varphi
\nonumber \\
&-&\frac{\tau^4}{252}\left(\frac{10}{3} \varphi^{,ik}\varphi_{,ik}-
\left(\nabla^2 \varphi\right)^2 \right)\, .  
\end{eqnarray}
The energy constraint also provides the solution for the scalar perturbation 
of the traceless part of the second-order spatial metric 
$\chi^{(2)}_{ij}$, while 
the vector and tensor parts follow by solving the momentum constraint and the 
remaining evolution equation, according to the same procedure outlined in 
Ref.~\cite{MMB}. We find $\chi^{(2)}_{ij}=-\frac{20}{9}
(1-a_{\rm nl})\tau^2 \varphi\varphi_{,ij}
-\frac{20}{9}\left( \frac{3}{2}-a_{\rm nl}  \right)\tau^2 
\varphi_{,i}\varphi_{,j} 
+\frac{20}{27}\tau^2\left[(1-a_{\rm nl})\varphi\nabla^2\varphi+
\Big( \frac{3}{2}-a_{\rm nl}\Big) 
\varphi^{,k}\varphi_{,k}\right]\delta_{ij}+ 
\frac{19}{126} \tau^4 \varphi^{,k}_{~,i}\varphi_{,kj}
+\frac{\tau^4}{126} \left( - 12\varphi_{,ij}
\nabla^2\varphi+4 \Big(\nabla^2\varphi \right)^2 \delta_{ij} 
-\frac{19}{3}\varphi^{,kl}\varphi_{,kl} \delta_{ij} \Big)+
\Delta^{(2)}_{ij}$, where $\Delta^{(2)}_{ij}$, describing second-order 
tensor modes 
generated by  
linear scalar perturbations and possible time-independent terms arising
from the initial conditions,  will not be necessary
for our purposes. 
Finally,  for the density contrast in the synchronous gauge 
we plug  the solution~({\ref{psi2s}) into Eq.~(\ref{delta2s}) to obtain
\begin{eqnarray}
\label{solution:delta2s}
\delta^{(2)}&=&
\frac{\tau^4}{126} \left[ 5 \left( \nabla^2 \varphi \right)^2 +
2 \varphi^{,ij}\varphi_{,ij}\right] +
\frac{10}{9}\tau^2\left[\left( \frac{3}{4}-a_{\rm nl}\right) \varphi_{,k}
\varphi^{,k} \right. \nonumber \\
&+& \left. \left(2-a_{\rm nl}\right)\varphi \nabla^2\varphi \right]\, .
\end{eqnarray}
The expression 
obtained for the metric perturbation $\psi^{(2)}$ is characterized by a 
large-scale part $(\tau \rightarrow 0)$  which is 
dominated by the primordial NG contribution. It is such  
 primordial NG which propagates onto smaller scales, 
once the mode reenters  the horizon during the matter-dominated epoch.
This is  evident 
from the dependence on $a_{\rm nl}$ of the term proportional to $\tau^2$ in 
Eq.~(\ref{psi2s}).  

The last step  of our computation consists in 
expressing the relevant quantities in the Poisson gauge~\cite{Bert}, 
{\it i.e.} the generalization beyond linear order of the longitudinal gauge, 
by which  a more direct comparison with the standard Newtonian 
approximation adopted in LSS observations and N-body simulations is  possible.
Second-order coordinate tranformations may be written as  
$\tilde{x}^\mu=x^\mu-\xi^\mu_{(1)}-(\xi^\mu_{(2)}-\xi^\mu_{(1),\nu}
\xi^\nu_{(1)})/2$, with  $\xi^0_{(r)}=\alpha^{(r)}$ and 
the space shift splits into the scalar and vector parts as 
$\xi^i _{(r)}=\beta^{,i}_{(r)}+d^i_{(r)}$. 
The density contrast and the velocity perturbation
transform  correspondingly as in Ref.~\cite{MMB}. 
The first-order 
transformation parameters to go from the
synchronous to the Poisson gauge are given by~\cite{MMB} 
$\alpha^{(1)}=\tau\varphi/3$, $\beta^{(1)}=\tau^2\varphi/6$ and $d^{(1)i}=0$. 
It follows that at linear order the metric perturbations are given 
in terms of the peculiar gravitational potential $\varphi$~\cite{MMB} as
$\psi^{(1)}=\phi^{(1)}=\varphi$ where $\phi=\phi^{(1)}+\phi^{(2)}/2$ and
$\delta g_{00}=-a^2(1+2\phi)$ defines the lapse function in the Poisson gauge. 
Linear tensor perturbations are gauge-invariant while 
the density contrast and the velocity perturbations read~\cite{MMB} 
$\delta^{(1)}=-2\varphi+\tau^2\nabla^2\varphi/6$ and 
$v^{(1)i}=-\tau\varphi^{,i}/3$ respectively. 
For simplicity we do not report here the explicit expressions we find for the 
second-order transformation parameters and we give the results for the 
relevant perturbations in the Poisson gauge. We refer the reader 
to Eqs.~(5.12),~(5.13) and~(5.14) of Ref.~\cite{MMB}, 
which give the general formulae for $\beta^{(2)}$, $d^{(2)}_i$ 
and $\alpha^{(2)}$. We just mention here that 
the initial conditions for the second-order perturbations  
will show up in such a computation. 
For the density constrast and the peculiar velocity
in the Poisson gauge we find  
\begin{eqnarray}
\label{delta2P}
\delta_P^{(2)}&=&\frac{\tau^4}{126}\left[5\left( \nabla^2 \varphi\right)^2 
+2 \varphi^{,ij}\varphi_{,ij}+7\varphi^{,i}\nabla^2\varphi_{,i} \right] 
+\frac{\tau^2}{9}\bigg[10\left(\frac{9}{20}-
a_{\rm nl}\right)\varphi_{,k}\varphi^{,k} 
\nonumber \\
&+&10\left( \frac{8}{5}-a_{\rm nl} \right)\varphi \nabla^2\varphi
+\frac{36}{7} \Psi\bigg]
+\frac{20}{3} \left(a_{\rm nl}-\frac{2}{5}\right)\varphi^2-24\Theta\, , \\
\label{v2P}
v_P^{(2)i}&=&\frac{\tau^3}{9}\left(-\varphi^{,ij}\varphi_{,j}+\frac{6}{7} 
\Psi^{,i}\right)-2\tau\bigg[\frac{10}{9}\left( \frac{21}{20}-a_{\rm nl}
 \right) 
\varphi \varphi^{,i}
+2\Theta^{,i} \bigg] \nonumber \\
&+&\frac{4\tau}{9}\varphi \varphi^{,i}-d^{i'}_{(2)}\, ,
\end{eqnarray}
where $\nabla^2 \Theta\equiv\Psi-\frac{1}{3} \varphi^{,i}\varphi_{,i}$ with 
$\nabla^2 \Psi\equiv-\frac{1}{2}[(\nabla^2 \varphi)^2-\varphi_{,ik}
\varphi^{,ik}]$. Eqs.~(\ref{delta2P}) and ~(\ref{v2P}) are the main results of 
this {\it Letter}. In  particular they clearly show how the primordial NG 
which is initially generated 
on large scales is transferred to the density contrast and
peculiar velocity
on subhorizon scales. The expression for 
the density contrast is made of three contributions: the standard second-order
Newtonian piece (proportional to $\tau^4$) which is insensitive
to the non-linearities in the initial conditions; a Post-Newtonian (PN)
piece  (proportional to $\tau^2$) which carries the most relevant
information on primordial NG; the super-horizon terms (independent of
$\tau$).  Our findings show in a clear way that the information on the
primordial NG set on super-Hubble scales flows into the PN terms, leaving
an observable imprint in the LSS.

It might also be  useful to write the density contrast in Fourier space in 
terms of the linear density contrast, by 
defining a kernel ${\cal K}({\bf k}_1,{\bf k}_2;\tau)$, depending on the 
wavevector of the perturbation modes, as 
\begin{eqnarray}
&&\delta_{\bf k}(\tau)=
\delta^{(1)}_{\bf k}(\tau)
+\int \frac{d{\bf k}_1 d{\bf k}_2}{(2\pi)^3} 
{\cal K}({\bf k}_1,{\bf k}_2;\tau)\delta^{(1)}_{{\bf k}_1}
\delta^{(1)}_{{\bf k}_2}
\delta({\bf k}_1+{\bf k}_2-{\bf k}) \, .
\end{eqnarray}       
The kernel ${\cal K}$ reads 
\begin{eqnarray}
\label{KER}
{\cal K}({\bf k}_1,{\bf k}_2;\tau)&=&
\frac{5}{7}+\frac{2}{7}\frac{
\left({\bf k}_1 \cdot {\bf k}_2\right)^2}{k_1^2 k_2^2}+\frac{1}{2} 
\frac{\left({\bf k}_1 \cdot {\bf k}_2\right)}{k_1^2 k_2^2} 
\left(k_1^2+k_2^2\right) \nonumber \\
&-&6 f_{\rm nl}({\bf k}_1,{\bf k}_2;\tau)\frac{k^2}{k_1^2 k_2^2\tau^2}\, ,
\end{eqnarray}
where
\begin{eqnarray}
f_{\rm nl}({\bf k}_1,{\bf k}_2;\tau)&=&
-\frac{5}{3}\, a_{\rm nl}
+\frac{63({\bf k}_1 \cdot {\bf k}_2)+172(k_1^2+k_2^2)}{42 k^2}
\nonumber\\ 
&+&\frac{(k_1^2+k_2^2)({\bf k}_1 \cdot {\bf k}_2)}{k_1^2 k_2^2 k^2}
\times \left(\frac{4}{7}({\bf k}_1 \cdot {\bf k}_2)
+k_1^2+k_2^2\right)\nonumber \\
&-&\frac{6}{7}\frac{k_1^2k_2^2}{k^4}+\frac{6}{7}
\frac{({\bf k}_1 \cdot {\bf k}_2)^2}{k^4}\, .
\label{kernel}
\end{eqnarray}
\begin{figure}
\centering
\includegraphics{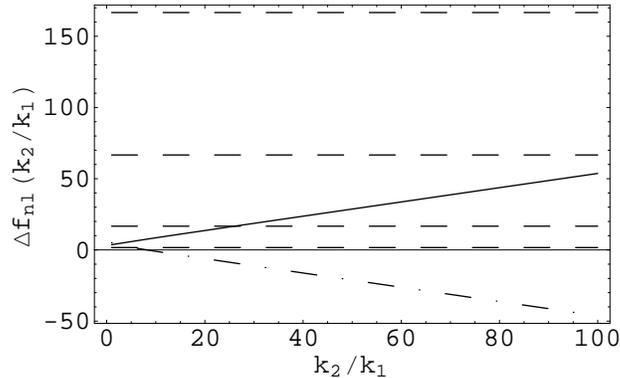}
\caption{Wavenumber dependence of the non-linearity parameter 
$f_{\rm nl}$. The continuous and dashed-dotted lines correspond to the 
k-dependent part $\Delta f_{\rm nl}$ of $f_{\rm nl}$ ($f_{\rm nl}$ for 
$a_{\rm nl}=0$), 
with an angle between ${\bf k}_1$ and 
${\bf k}_2$ of $\pi/3$ and $2\pi/3$ respectively. The horizontal lines show 
the k-independent part $|5 a_{\rm nl}/3|$ for $a_{\rm nl}=1, 10, 40, 100$. }
\label{plot}
\end{figure}
In order to obtain the expression in Eq.~(\ref{KER}) we have performed an 
expansion in $(k_{1,2} \tau)^{-1} \ll 1$ up to terms $(k_i \tau)^{-2}$ 
starting from Eq.~(\ref{delta2P}). 
The non-linearity paramater $f_{\rm nl}$ defined in this
way generalizes the standard definition of Refs. \cite{verde}
inferred from the Newtonian gravitational potential $\Phi=\varphi+
f_{\rm nl}\left(\varphi^2-\langle \varphi^2\rangle\right)$. Our findings 
indicate that the non-Gaussianity in the density contrast shows a non-trivial
shape dependence which might help in detecting  the PN nonlinearities. 
In Fig.~\ref{plot} we plot the k-dependence of the non-linearity parameter 
$f_{\rm nl}$ which can be useful  
in distinguishing the different contributions in the measured bispectrum.
The mass-density bispectrum is given by $\langle 
\delta({\bf k}_1) \delta({\bf k}_2) \delta({\bf k}_3)
\rangle= (2 \pi)^3 \delta^{(3)}({\bf k}_1+{\bf k}_2+{\bf k}_3) 
[2 {\cal K}({\bf k}_1,{\bf k}_2;\tau) P_\delta(k_1) P_\delta(k_2) + 
{\rm cycl.}]$, where $P_\delta(k)$ is the linear 
mass-density power spectrum. 
By inspecting  Eq.~(\ref{kernel}) 
we can   roughly estimate  the relevance of the primordial NG 
contribution to the mass-density bispectrum (proportional to
$a_{\rm nl} (1/\tau k)^2$)
with respect to the 
Newtonian one. 
For the two to be comparable on a given scale $\lambda \sim k^{-1}$,
reminding the reader that $f_{\rm nl}\sim- \frac{5}{3}a_{\rm nl}$ 
($|a_{\rm nl}|\gg 1$), 
we 
find that the non-linearity parameter $f_{\rm nl}$ must be as large as
$f_{\rm nl}\sim (10^5/1+z)\left({\rm Mpc}/\lambda\right)^2$. 
Thus, we confirm that, 
on a comoving scale  $\lambda \sim {\rm few}\, {\rm Mpc}$, 
a value of the non-linearity parameter 
$f_{\rm nl} \sim 10^3$ 
is required for the primordial NG to leave observable
effects in the clustering of galaxies at low redshift \cite{verde}.
Going to higher redshift would allow to pin down lower 
values of $f_{\rm nl}$ \cite{scocci}, provided  
the galaxy-to-dark matter bias evolution is accurately handled \cite{mat}, 
or that observables directly related to the dark matter density contrast are 
considered (see e.g. Ref.~\cite{cooray}). As far as the velocity divergence
is concerned, in the limit of large NG, we also recover the standard
Newtonian formula relating the NG in the peculiar velocity to the one present
in the gravitational potential $\Phi$, $\nabla\cdot v_P=-(2/3{\cal H})\nabla^2
\Phi$. Finally, one can easily extend our findings for large NG to the case 
in which dark energy is present. For the density constrast, it suffices 
to replace, in the term proportional to $a_{\rm nl}$, 
$\tau^2$ with $4 D_+^2 {\cal H}_0^{-2}
\Omega_{0m}^{-1}$, where $D_+$ is the standard Newtonian linear
 growth factor of density
perturbations \cite{ct} and $\Omega_{0m}$ is the presenty-day matter density
parameter. As for the peculiar velocity, one has to replace,   
in the term proportional to $a_{\rm nl}$, 
$\tau$ with $2D_+{\cal H}_0^{-2}{\cal H}f(\Omega_m)$, where $f(\Omega_m)=
d{\rm ln}~D_+/d{\rm ln}~ a$ \cite{ct}.

\section*{Acknowledgments}  
We would like to thank E. Branchini and C. Porciani for useful discussions.

\end{document}